\documentstyle[aps,pre,twocolumn,floats,epsfig]{revtex}

\begin{document}

\draft
\wideabs{

\title{Breaking conjugate pairing in thermostatted billiards 
                 by magnetic field}

\author{M. Dolowschi\'ak and Z. Kov\'acs}

\address{Institute for Theoretical Physics, E\"otv\"os University\\
          Pf.\ 32, H--1518 Budapest, Hungary}

\maketitle

\begin{abstract} 
We demonstrate that in the thermostatted three-dimensional 
Lorentz gas the symmetry of the Lyapunov spectrum can be broken 
by adding to the system an external magnetic field not perpendicular
to the electric field.
For perpendicular field vectors, there is a Hamiltonian reformulation
of the dynamics and the conjugate pairing rule still holds.
This indicates that symmetric Lyapunov spectra has nothing to do with
time reversal symmetry or reversibility; instead, it seems to 
be related to the existence of a Hamiltonian connection.
\end{abstract}

}    

\section{Introduction}

Thermostatted dynamical systems have raised considerable interest 
recently as a testing ground for ideas in nonequilibrium 
statistical mechanics \cite{thermo}. 
In particular, questions concerning the role played by
chaotic dynamics in the appearance of nonequilibrium stationary
states in dissipative systems have been in the focus of research
activities \cite{nonequi}. 
One of the most remarkable features of these models is that they are 
dissipative {\em and} time reversal symmetric at the same time.
Some, but not all, thermostatted systems have another interesting
common property known as the {\em conjugate pairing rule} (CPR):
the Lyapunov exponents of the system form
pairs summing up to the same (negative) value \cite{CPR}.
It has a practical relevance too: with CPR, one pair of Lyapunov 
exponents can be used to determine the sum of {\em all} the exponents, 
which is known to be connected to the transport properties of the 
system \cite{Lyap-sum}.
CPR is trivially present in conservative Hamiltonian
systems (the sum being zero due to simplecticity); 
however, there is no obvious reason to expect anything
similar in dissipative systems.
In fact, CPR in thermostatted systems was first discovered by numerical 
studies \cite{CPR}.

The simplest system in which CPR can be checked is the three-dimensional
(periodic) Lorentz gas (3DLG): due to its three degrees of freedom,
it has four nontrivial Lyapunov exponents.
Dettmann {\it et al.} have shown numerically 
\cite{DettMorrRond:3DLG} 
that the 3DLG with external electric field and
Gaussian isokinetic (GIK) thermostat  
exhibits conjugate pairing;
later, this has been proven analytically 
for conservative forces and hard-wall scatterers 
\cite{CPR-proof}.
It has also been demonstrated \cite{DettMorr:Ham} that 
this system can be connected to a Hamiltonian dynamics.

In this paper, we check the effect of an external magnetic field
on the validity of CPR in the GIK thermostatted cubic lattice 3DLG.
In particular, we will focus on two features possibly related
to CPR: reversibility (an extension of time reversal symmetry) 
and the existence of a Hamiltonian formulation.
Both can be controlled by the direction of the magnetic field
with respect to that of the electric field and the lattice.
Our numerical results show that CPR is not affected by breaking 
reversibility, and it also holds for cases with perpendicular
electric and magnetic field vectors for which there is a
connection to Hamiltonian dynamics.
However, CPR breaks down for nonperpendicular fields, i.e.\ 
in the case when no Hamiltonian connection has
been found.

In Sec.~\ref{sec-bill}, the equations of motion for
billiards with GIK thermostat in magnetic field are presented,
together with a discussion of reversibility and the Hamiltonian
connection for perpendicular fields.
The numerical results for the 3DLG and our conclusions
are presented in Sec.~\ref{sec-num} and  \ref{sec-concl},
respectively.

\section{Thermostatted billiards in magnetic field}
\label{sec-bill}

\subsection{The dynamics}
\label{sec-dyn}

The kinetic energy of a particle moving under the influence of 
external fields can be kept constant by adding a special 
frictionlike force to the system.
Since this force can be deduced from Gauss's principle of least
constraint, and it is kinetic energy that is kept constant, the technique
is called Gaussian isokinetic (GIK) thermostat \cite{thermo}.
In billiards, this is equivalent of particle momentum ${\bf p}$ 
changing only in direction, but not in magnitude $p$, during the
``free'' flights between collisions with the hard-wall boundaries.
Choosing the unit of mass to be the mass of the particle, 
the corresponding  equations of motion are
\begin{equation}
\label{eq-GIK}
   \dot{\bf q} = {\bf p}, \ \ 
   \dot{\bf p} = {\bf F}_e - \alpha {\bf p}
\end{equation}
where ${\bf q} = (x,y,z)$ is the position of the particle, ${\bf F}_e$ 
stands for the external forces, 
while the GIK thermostat corresponds to the choice
\begin{equation}
   \alpha  =  \frac{{\bf F}_e {\bf p}}{p^2}.
\label{eq-alfa}
\end{equation}
For simplicity, we will choose length and time units in our studies 
so that $p=1$,
but care must be taken when substituting 1 for $p^2$ in terms like 
$\alpha$ above,
especially in the derivation of tangent space equations 
for the calculation of Lyapunov exponents.

In our model, the external force ${\bf F}_e$ contains the (constant)
electric and magnetic fields ${\bf E}$ and ${\bf B}$:
\begin{equation}
\label{eq-force}
  {\bf F}_e = {\bf E} + {\bf p} \times {\bf B}
\end{equation}
(we have defined the unit of electric charge to be that of the particle).
The full dynamics also includes the secular collisions
with the hard-wall boundaries, changing the momentum ${\bf p}_i$ 
to ${\bf p}_f$ instantaneously:
\begin{equation}
  {\bf p}_f = ( I - 2 {\bf n} \circ {\bf n}) {\bf p}_i 
\end{equation}
where $I$ is the ($3 \times 3$) identity matrix, 
${\bf n}$ is the normal vector of the boundary at the collision 
point and `$\circ$' denotes the diadic product.
In our 3DLG, the scatterers are hard spheres 
of radius $R$,
arranged into a regular cubic lattice with distance $d$ between
the centers of nearest neighbor scatterers.
For simplicity, we choose the length scale so that $R=1$.

\subsection{Reversibility}
\label{sec-rev}

Without magnetic field, Eqs.~(\ref{eq-GIK}) and (\ref{eq-alfa}) 
ensure time reversal symmetry for the dynamics, which means that
for each solution 
$\Gamma_+ (t) = ({\bf q}(t),{\bf p}(t))^T$
there exists another one tracing the same path backward in time: 
\begin{equation}
\label{eq-time-rev}
 \Gamma_- (t) = ({\bf q}(-t),-{\bf p}(-t))^T.
\end{equation}
The pairing of solutions by time reversal symmetry is important
in these models: it is used e.g.\ in showing that the average
current flows in the direction of the external electric 
field \cite{current}. 
This symmetry cannot hold if ${\bf B} \neq 0$, but the more
general property of {\em reversibility} \cite{LambRob} 
may still be true,
depending on the particular choice of ${\bf E}$ and ${\bf B}$.
Reversibility means the existence of a transformation
$G$ in phase space which is an involution 
(i.e., $G^2$ is the identity) 
mapping each solution $\Gamma_+ (t)$ to another one $\Gamma_- (t)$ 
in the following manner:
\begin{equation}
   \Gamma_- (t) = G \Gamma_+ (-t). 
\end{equation}
In terms of the the phase space flow $\phi^t$ defined by
$\Gamma (t) = \phi^t \Gamma (0)$,
this requirement can be written as
\begin{equation}
\label{eq-flowrev}
    G \phi^t G = \phi^{-t} \ ,
\end{equation}
i.e., bracketing the flow by $G$ ``reverses the direction of time''.

Ordinary time reversal symmetry is equivalent to $G=G_0$ just flippping
the direction of the momentum: 
$G_0 ({\bf q},{\bf p}) = ({\bf q},-{\bf p})^T$.
For ${\bf B} \neq 0$, the flow can be reversed by the transformation 
$G_B = M G_0$ where $M$ is
a mirroring of ${\bf q}$ and ${\bf p}$ with respect to the plane 
containing ${\bf E}$ and ${\bf B}$ (the proof of this statement is left
to the Appendix).
In the Lorentz gas, reversibility of the full dynamics also requires that
the invariant plane of $M$ be a symmetry plane of the lattice too.
This gives us an easy way to control reversibility in the Lorentz gas:
choosing directions for ${\bf E}$ and ${\bf B}$ in a symmetry
plane of the lattice leads to reversible dynamics, otherwise we have
no reversibility.

\subsection{Hamiltonian formalism}
\label{sec-Ham}

A nontrivial result for GIK thermostatted systems without magnetic field 
is that a Hamiltonian formulation of the dynamics exists  
provided the force ${\bf F}_e$ is the gradient of a scalar field 
$-\Phi({\bf q})$ \cite{DettMorr:Ham}.
Then there is a Hamiltonian $H({\bf Q}, {\bf P})$ 
so that the GIK equations of motion for the physical variables 
${\bf q}$ and ${\bf p}$ can be obtained from 
the canonical equations of motion for ${\bf Q}$ and ${\bf P}$
through a suitable coordinate transformation.
It is straightforward to check that the Hamiltonian
$H ({\bf Q}, {\bf P}) = \frac{1}{2} [e^{\Phi} {\bf P}^2 - e^{-\Phi}]$
has canonical equations leading to Eq.~(\ref{eq-GIK}) 
if one assumes the transformations 
${\bf q} = {\bf Q}$ and ${\bf p} = e^{\Phi} {\bf P}$.
However, it is important to stress that this connection holds 
only if we make explicit use of the constraint 
$p=1$ and its equivalent $H=0$ in the GIK and canonical
equations, respectively.

The extension of the Hamiltonian formulation to cases with 
${\bf B} \neq 0$ is not as obvious as for conservative
systems because of the factor $e^{\Phi}$ in front of ${\bf P}^2$
in the ``kinetic energy'' term of the Hamiltonian.
Nevertheless, we may still follow a similar route 
by defining $\Phi({\bf q})$ through ${\bf E} = - \nabla \Phi$ as usual
and replacing ${\bf P}$ by ${\bf P} - {\bf a}({\bf q})$ in $H$,
where the vector ${\bf a}({\bf q})$ is connected to the magnetic field.
This leads to the Hamiltonian
\begin{equation}
\label{eq-HamB}
    H_B ({\bf Q}, {\bf P}) = 
          \frac{1}{2} [e^{\Phi} ({\bf P} - {\bf a})^2 - e^{-\Phi}].
\end{equation}
A lengthy but straightforward calculation shows \cite{KZDM}
that the canonical equations for $H_B = 0$ can be connected
to the GIK equations of motion for $p=1$ by the transformation
\begin{equation}
\label{eq-qQpP}
      {\bf q} = {\bf Q}, \ \ \
      {\bf p} = e^{\Phi} ({\bf P} - {\bf a})
\end{equation}
if we assume the following relationship between ${\bf B}$ and ${\bf a}$:
\begin{equation}
\label{eq-Ba}
        {\bf B} =  e^{\Phi} \mbox{rot}\ {\bf a}. 
\end{equation}
Note that this is an extension of the usual relationship 
${\bf B} =  \mbox{rot}\ {\bf a}$ for the GIK thermostat.
However, due to the presence of $e^{\Phi}$ in Eq.~(\ref{eq-Ba}),
we do not necessarily have a solution ${\bf a}$
for arbitrary ${\bf E}$ and ${\bf B}$.
Indeed, since  ${\bf B}$ must satisfy Maxwell's equation  
$\mbox{div}\ {\bf B} = 0$,
this condition leads to the restriction ${\bf E} {\bf B} = 0$.
Therefore, we can use the Hamiltonian formulation given above 
only in the case when ${\bf B}$ is perpendicular to ${\bf E}$.

\section{Numerical results}
\label{sec-num} 

\begin{figure}
\epsfig{file=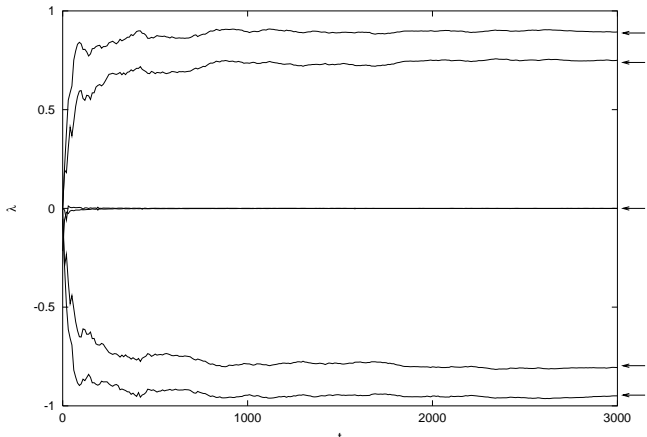,width=8.5cm}
\caption{%
Time evolution of the measured Lyapunov exponents in the 3DLG. 
The scatterers are balls with radius $R=1$ arranged in a cubic
lattice with lattice constant $d=2.3$.
The external fields are ${\bf E} = (0.3, 0, 0)$ and 
${\bf B} = (0, 0.5 \cos \phi, 0.5 \sin \phi)$ with $\phi = \pi/20$.
The arrows show the long-time values of the exponents ($t=10^8$);
$\lambda_3$ and $\lambda_4$ converge to 0 as expected.}
\label{fig1}
\end{figure}

We have calculated numerically 
the Ljapunov spectrum of the GIK thermostatted
3DLG with constant electric and magnetic fields.  
The Lyapunov exponents $\lambda_1 > \lambda_2 > \ldots > \lambda_6$
can be measured by following the evolution of a
full set of linearly independent tangent space vectors along 
a very long trajectory and applying repeated 
reorthogonalization and rescaling to them; 
see Ref.~\cite{Benettin} for a detailed
description of this method. 
The effect of collisions were taken into account by the formula 
presented  in Ref. \cite{Posch:coll}.
In all cases studied we have obtained finite time exponents
converging to their infinite time limits as in the example 
plotted in Fig.~\ref{fig1}. 
The fluctuations in the measured values typically tend to 
zero as $1/\sqrt{N}$,
where $N$ is the number of collisions,
so for reliable results we needed very long runs with
$N=10^7$ collisions or more.
The data also show that the largest Ljapunov exponent is 
positive, i.e.\ the motion is chaotic,
and that two of the exponents are zero as expected.

We have choosen the coordinate axes $x$, $y$ and $z$  
aligned with the lattice axes.
Through the directions of the field vectors, we can have 
reversible or non-reversible dynamics in our model,
with or without a Hamiltonian representation, independently.
In the simulations, we fixed ${\bf E}$ along the $x$ axis,
so that it lies in the symmetry planes $y=0$ and $z=0$,   
and controlled the above properties by chosing the direction
of ${\bf B}$ accordingly. 
In particular, the dynamics is reversible e.g.\ for $B_y =0$;
meanwhile, there exists a Hamiltonian formulation
as given in Sec.~\ref{sec-Ham} for $B_x = 0$.

\begin{figure}
\epsfig{file=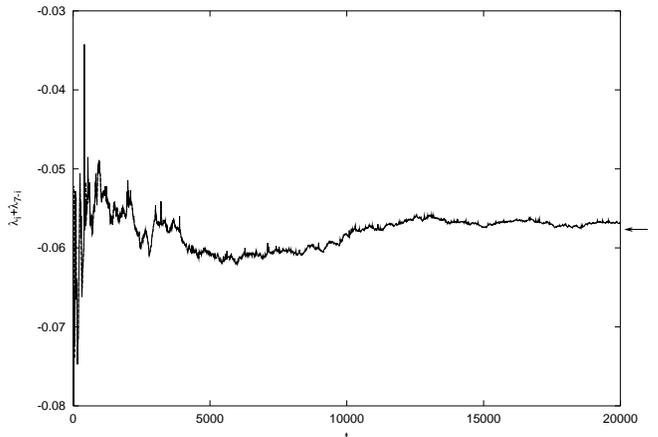,width=8.5cm}
\caption{%
Time evolution of the sums $\lambda_1 + \lambda_6$
(solid line)
and $\lambda_2 + \lambda_5$ (dashed line),
as obtained from the data in Fig.~\ref{fig1}.
The two graphs are practically indistinguishable for 
$t > 5000$.
The arrow shows the asymptotic value of the sums ($t=10^8$).}
\label{fig2}
\end{figure}

Figure~\ref{fig1} shows the results of a simulation for
${\bf B} = (0, B \cos \phi, B \sin \phi)$ with $\phi = \pi /20$, 
i.e., for perpendicular fields and {\em without} reversibility.
In Fig.~\ref{fig2}, we plotted the sums of the pairs
$\lambda_1 + \lambda_6$ and $\lambda_2 + \lambda_5$.
They both converge rapidly to the same value, thus
CPR seems to hold in this case.
It is also worth noting that the difference between the two sums
disappears much faster than the fluctuations in the
individual exponents.
We have obtained similar results for other values of the
angle $\phi$, including reversible flows (e.g. $\phi = 0$).
These results demonstrate that reversibility is not needed for 
CPR to hold.

\begin{figure}
\epsfig{file=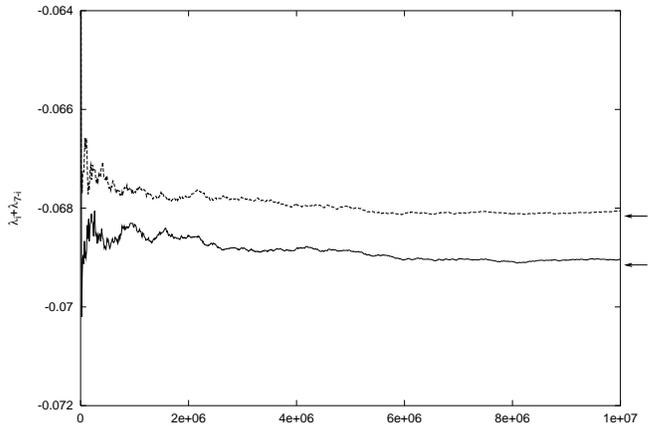,width=8.5cm}
\caption{Same as Fig.\ 2 for nonperpendicular fields;
${\bf B} = ( 0.5 \sin \phi, 0, 0.5 \cos \phi)$
with $\phi = 7\pi/20$.
Notice the change of scales with respect to Fig.\ 2.}
\label{fig3}
\end{figure}

In the second type of simulations 
we have choosen ${\bf B} = ( B \sin \phi , 0 , B \cos \phi)$,
so that for $\phi \neq 0$ the two field vectors are not
perpendicular and the Hamiltonian formulation of Sec.~\ref{sec-Ham}
does not apply.
The numerical Lyapunov spectrum looks qualitatively the same
as in Fig.~\ref{fig1}, 
but the sums of the two pairs seem to converge to different
values as shown in Fig.~\ref{fig3} for the angle 
$\phi = 7\pi/20$. 
In other words, CPR is broken in this case;
other values of $\phi \neq 0$ have lead to similar results.
The difference between Figs.~\ref{fig2} and \ref{fig3}
is very clear: the quantity 
${\Delta=\lambda_1+\lambda_6-(\lambda_2+\lambda_5)}$ converges to
zero quite fast if CPR holds, while it stays 
definitely positive in the case without CPR.

\section{Conclusions}
\label{sec-concl}

We have demonstrated that in the GIK thermostatted 3DLG the
CPR can be broken by an external magnetic field which is not
perpendicular to the electric field.
For perpendicular fields, however, CPR holds, and 
the convergence of the pair sums to each other seems to be much 
faster than that of the individual exponents indicating 
that CPR is valid for all times in these cases 
just as in the 3DLG without magnetic field 
\cite{DettMorrRond:3DLG}.
This phenomenon is called {\em strong} CPR.
The perpendicular cases are also characterized by the
existence of a Hamiltonian formulation.
There exist other nontrivial examples for
systems with strong CPR and a Hamiltonian
formulation, too: e.g., the Gaussian isoenergetic thermostat
with a special interparticle potential \cite{Dett:GIE}
or the ideal Sllod gas \cite{Searl:Sllod,KZDM}.
These examples
suggest that there may be a direct connection between 
strong CPR and the existence of a Hamiltonian formulation.
We will examine this question in a separate paper \cite{KZDM}.
Although we are not aware of any counterexamples, 
the question concerning the existence of systems with 
strong CPR but without a Hamiltonian formulation is still open. 

Our results also show that time reversal symmetry,
or reversibility in general, is not needed for CPR to hold.
Indeed, one of the first examples for CPR in dissipative
systems has been a Hamiltonian system with a constant viscous
damping \cite{Dressler} which has no time reversal symmetry.

\section*{Acknowledgments}

This work was supported by the Bolyai J\'anos Research Grant of
the Hungarian Academy of Sciences and by the Hungarian Scientific 
Research Foundation (Grant Nos.\ OTKA F17166 and T032981).

\section*{Appendix} 

We show that the flow defined by
Eqs.~(\ref{eq-GIK})--(\ref{eq-force})
is reversible with respect to the transformation $G_B = M G_0$ as given
in Sec.~\ref{sec-rev}.
Equation~(\ref{eq-flowrev}) can be rewritten 
for the time derivative $F$ of the flow
as $G F G = -F $.
From Eq.~(\ref{eq-GIK})
one can see that $F ({\bf q},{\bf p}) = ({\bf p},{\bf f})^T$, 
whith ${\bf f}({\bf q},{\bf p})$ given by the expression for 
$\dot{\bf p}$.
Now we can write that
\begin{eqnarray*}
  G_B F G_B ({\bf q},{\bf p}) 
               & = & G_B F (M {\bf q},-M{\bf p}) \\
               & = & G_B (-M{\bf p},{\bf f}(M{\bf q},-M{\bf p})) \\
               & = & (-{\bf p},-M{\bf f}(M{\bf q},-M{\bf p}))^T  \\
               & = & (-{\bf p},-{\bf f}({\bf q},{\bf p})^T.  
\end{eqnarray*}
The last equality gives us the condition 
\begin{equation}
\label{eq-condf}
      {\bf f}({\bf q},{\bf p}) = M{\bf f}(M{\bf q},-M{\bf p})
\end{equation}
for the force acting on the particle.

Since ${\bf f}$ consists of the two parts of the
Lorentz force and the thermostat, we can check these
terms separately. 
For the electric field this means that ${\bf E} = M {\bf E}$,
i.e., ${\bf E}$ must be in the invariant plane of $M$.
For the term ${\bf p} \times {\bf B}$, the right hand side of
Eq.~(\ref{eq-condf})
reads as $M(-M{\bf p} \times {\bf B}) = 
  -M({\bf p}_{\parallel} \times {\bf B} - {\bf p}_{\perp} \times {\bf B}) =
  -M({\bf p}_{\parallel} \times {\bf B}) + M({\bf p}_{\perp} \times {\bf B})$,
where ${\bf p}_{\parallel}$ and ${\bf p}_{\perp}$ denote the components
of ${\bf p}$ parallel and perpendicular to the invariant plane of $M$,
respectively.
If ${\bf B}$ is in this plane, then 
$-M({\bf p}_{\parallel} \times {\bf B}) = {\bf p}_{\parallel} \times {\bf B}$
and $M({\bf p}_{\perp} \times {\bf B}) = {\bf p}_{\perp} \times {\bf B}$,
so the magnetic part of the Lorentz force also satisfies
Eq.~(\ref{eq-condf}).
As for the thermostatting force,
$M(({\bf E} \cdot M{\bf p})M{\bf p}) = (({\bf E} \cdot M{\bf p}) M^2 {\bf p} = 
({\bf E} \cdot {\bf p}) {\bf p}$ also holds if ${\bf E} = M{\bf E}$.
Thus the flow is reversed by $G_B = M G_0$ if the field
vectors ${\bf E}$ and ${\bf B}$ are invariant under $M$.


\end{document}